\documentstyle[aps,preprint]{revtex}

\def \cT {{\cal T}}

\def \cG {{\cal G}}
\def\cB{{\cal B}}

\def \cT {{\cal T}}

\def \cA {{\cal A}}

\title{Fano resonances and
Aharonov-Bohm  effects in transport through a square quantum dot
molecule}

\author{Z. Y. Zeng$^{1,2}$, F. Claro$^{1}$, and Alejandro \ P\'erez$^{1}$}
\address{1.Facultad de F\'isica, Pontificia Universidad
Cat\'olica de Chile, Casilla 306, Santiago 22, Chile \\
2. Department of Physics, Hunan Normal University, Changsha
410081, China and CCAST(World Laboratory). P. O. Box 8730, Beijing
100080, China}
\begin{document}
\date{\today}
\maketitle
\draft
\widetext

\begin{abstract}
We study the Aharonov-Bohm  effect in a coupled 2$\times$2 quantum
dot array with two-terminals. A striking conductance dip arising
from the Fano interference is found as the energy levels of the
intermediate dots are mismatched, which is lifted in the presence
of a magnetic flux. A novel five peak structure is observed in the
conductance for large mismatch. The Aharonov-Bohm  evolution of
the linear conductance strongly depends on the configuration of
dot levels and interdot and dot-lead coupling strengths. In
addition, the magnetic flux and asymmetry between dot-lead
couplings can induce the splitting and combination of the
conductance peak(s).

\end{abstract}


\section{Introduction}

The famous effect predicted by Aharonov and Bohm ({\bf AB}) in
1959 \cite{AB} concerns the essential influence of a vector
potential on the interference pattern of two beams of electrons
confined in multiply connected paths, over which the magnetic
field is zero. It is manifested by the periodic oscillation of the
conductance of a ring or a cylinder as a function of the enclosed
magnetic flux $\Phi$ \cite{Webb}. It has been shown that the {\bf
AB} effect also exists in some singly connected geometry such as a
point contact or a disk shape in a two-dimensional electron gas,
in which circulating edge states enclose a well-defined magnetic
flux \cite{Beenakker}. A significant hallmark in mesoscopic
experiments is the phase measurements of the transmission
amplitude through a quantum dot embedded in an {\bf AB} ring in
the Coulomb blockade regime \cite {Yacoby}. It revived interest on
the {\bf AB} effect in condensed matter physics \cite
{Hackenbroich}.

    Quantum dots are highly tunable artificial mesoscopic
structures \cite{Kouwenhoven}, usually called artificial atoms or
artificial molecules. Some novel physical properties, such as the
Coulomb blockade and the turnstile effect, have been demonstrated
in some elegant experiments \cite{Kouwenhoven}. At low
temperatures, transport through a single quantum dot is dominated
by resonant tunneling and Coulomb blockade. When the condition for
resonant tunneling is not accessible, co-tunneling events
involving the simultaneous tunneling of two or more electrons
through a virtual intermediate level can be dominant
\cite{Averin}. When quantum dots are connected in series with
tunneling barrier(s), it is expected that the coupling between
dots plays a crucial role in the transport properties of the
coupled structure. On the other hand, electron tunneling through
the coupled dot system is very sensitive to the incoherent
scattering, which has trivial effects on the transport properties
of a single dot \cite{Datta}. In the past decade, transport
through  coupled quantum dot systems has received considerable
amount of attention \cite{Haug,Stafford,Oosterkamp}.

 In this paper, we investigate coherent electronic transport
 through four coupled quantum dots located at the four
 corners of a square enclosing a magnetic flux. This kind
 of coupled dot structure is the basic unit of two-dimensional quantum
 dot arrays. Our goal is to find out how the conductance of such a
 structure depends on the coupling strength between dots,
 the arrangement of dot levels and the magnetic flux.
 In the present work, we do not take into consideration the
 interdot and intradot electron-electron interactions. We also
 assume that just one level is relevant on each dot.
 This allows us, on the one hand, to obtain
some exact analytical results. On the other hand, we can gain much
 more clear physical insight into the dependence of the conductance
of the system on the structure parameters (interdot couplings and
dot levels) and the magnetic flux. Though the model we consider is
simple, some novel and interesting results are obtained arising
from the interplay of the specific configuration of dot levels and
interdot couplings, and the magnetic flux.

The rest of the paper is organized  as follows. In Sec. II we
derive the current through the 2$\times$2 quantum dot array using
the Keldysh nonequilibrium Green's function formalism and a
recursive Green's function technique. Section III presents the
linear conductance spectra of the dot array in the symmetric and
asymmetric coupling case. We also investigate in detail the
Ahoronov-Bohm oscillation of the linear conductance in some
specific configurations of the dot levels and the interdot
couplings. Our concluding remarks are given in Sec. IV.

\section{Formulation}
We consider four quantum dots in a square array
 \cite {Yu} enclosing a magnetic flux $\Phi$ (Fig. 1), with two dots facing each other
 connected
to the left and right leads. For simplicity, we ignore the intra-
and inter-dot Coulomb interactions, and assume that just one
energy level is relevant at each dot. Then the hamiltonian
describing such a system is
\begin{eqnarray}
 H       &=& H_{Dot}+H_{Lead}+H_{Dot-Lead},\cr
 H_{Dot} &=& \sum\limits_{i=1}^{4}\epsilon_id_i^+
              d_i+(t_{12}d^+_1d_2+t_{13}e^{i\phi}d^+_1d_3
              +t_{24}d^+_4d_2+t_{34}d^+_4d_3+h.c.), \cr
 H_{Lead}&=& \sum\limits_{k\in L}\epsilon_{k}a^+_{k}a_{k}+\sum\limits_{p \in
              R}\epsilon_{p}b^+_{p}b_{p}, \cr
 H_{Dot-Lead}&=& \sum\limits_{k\in L}(V_{k1}a^+_kd_1+V_{k1}^*d^+_1a_k)+ \sum\limits_{p\in
                 R}(W_{p4}b^+_pd_4+W_{p4}^*d^+_4b_p),
\end{eqnarray}
where $\epsilon_i$ is the energy level of dot $i$,
$a_k$($a_k^+$),$b_p$($b_p^+$) and $d_i$($d^+_i$) ($i=1,2,3,4$) are
the annihilation (creation) operators of electrons in the left(L)
lead, right(R) lead and quantum dot $i$, and $t_{ij}$ and $V_{k1}$
($W_{p4}$) are the inter-dot and dot-lead coupling matrix
elements, respectively.  In Eq. $(1)$, a factor $e^{i\phi}$
($\phi=2\pi \Phi/ \Phi_0$, $\Phi_0=h/e$)is attached to $t_{13}$ to
account for the magnetic flux $\Phi$ through the structure.

The current flowing from the left lead to the structure is \cite
{Meir}
\begin{equation}
J_L=\frac{2ie}{\hbar}\int\frac{d\epsilon}{2\pi}\Gamma_L(\epsilon)\{f_L(\epsilon)
[G^r_{d_1 d_1}(\epsilon)-G^a_{d_1d_1}(\epsilon)]+ G^<_{d_1
d_1}(\epsilon)\}.
\end{equation}
The notation we use throughout is  $G^{r,a}_{XY}(t,t')=\mp
i\theta(\pm t\mp t') <\{ X(t),Y^+(t')\}>$,
$G^<_{XY}(t,t')=i<Y^+(t')X(t)>$, $X,Y=a_k,b_P, d_i$ $(i=1,2,3,4)$,
$\Gamma_L(\epsilon)=2\pi|V_{k1}|^2\delta(\epsilon-\epsilon_{k})$,
while $f_L(\epsilon)=[1+e^{(\epsilon-\mu_L)/k_BT}]^{-1}$ is the
Fermi-Dirac distribution function of the left lead, with chemical
potential $\mu_L$. The factor of 2 in Eq. (2) is due to the spin
degeneracy.

Following the recursive decoupling technique developed in Ref.
\cite {Zeng}, the retarded Green's function appearing in Eq. (2)
$G^r_{d_1d_1}(\epsilon)$ can be calculated from Dyson's equation
\begin{eqnarray}
G^r_{d_1d_1} &=&[\epsilon-\epsilon_1-\Sigma^r_{d_1}]^{-1}, \\
\Sigma^r_{d_1}&=& -\frac{i}{2}\Gamma_L+|t_{12}|^2 \tilde
       G^r_{d_2d_2}+|t_{13}|^2 \tilde
            G^r_{d_3d_3}+t_{12}t^*_{13}e^{-i\phi} \tilde
               G^r_{d_2d_3}+t_{12}^*t_{13}e^{i\phi} \tilde
               G^r_{d_3d_2}.
\end{eqnarray}
Here and in what follows $\tilde G^{r,a,<}_{d_id_j}$ denote the
Green's function which is decoupled from the quantum dot $k$
($k=min(i,j)-1$). In the above and following equations we drop the
argument $\epsilon$, and recover it wherever necessary. Following
a similar recursive decoupling method, one has
\begin{eqnarray}
\tilde G^r_{d_2d_2}&=&
\frac{(\epsilon-\epsilon_3)(\epsilon-\epsilon_4+
\frac{i}{2}\Gamma_R)-t_{34}t^*_{34}}{(\epsilon-\epsilon_2)(\epsilon-\epsilon_3)
(\epsilon-\epsilon_4+ \frac{i}{2}\Gamma_R)-
t_{24}t^*_{24}(\epsilon-\epsilon_3)-t_{34}t^*_{34}(\epsilon-\epsilon_2)},
\\
 \tilde G^r_{d_3d_3}&=&
\frac{(\epsilon-\epsilon_2)(\epsilon-\epsilon_4+
\frac{i}{2}\Gamma_R)-t_{24}t^*_{24}}{
(\epsilon-\epsilon_2)(\epsilon-\epsilon_3)(\epsilon-\epsilon_4+
\frac{i}{2}\Gamma_R)-
t_{24}t^*_{24}(\epsilon-\epsilon_3)-t_{34}t^*_{34}(\epsilon-\epsilon_2)},
\\
 \tilde G^r_{d_2d_3}&=&
\frac{t_{24}t^*_{34}}{
(\epsilon-\epsilon_2)(\epsilon-\epsilon_3)(\epsilon-\epsilon_4+
\frac{i}{2}\Gamma_R)-
t_{24}t^*_{24}(\epsilon-\epsilon_3)-t_{34}t^*_{34}(\epsilon-\epsilon_2)},
\\
 \tilde G^r_{d_3d_2}&=& \frac{t_{34}t^*_{24}}{
(\epsilon-\epsilon_2)(\epsilon-\epsilon_3)(\epsilon-\epsilon_4+
\frac{i}{2}\Gamma_R)-
t_{24}t^*_{24}(\epsilon-\epsilon_3)-t_{34}t^*_{34}(\epsilon-\epsilon_2)},
\end{eqnarray}
where
$\Gamma_R(\epsilon)=2\pi|W_{p4}|^2\delta(\epsilon-\epsilon_{p})$.
In the following we will assume that the coupling matrix elements
($V_{k1}$, $W_{p4}$ and $t_{ij}$) are real since the tunnel rates
just depend on the amplitude of the coupling matrix elements.
Substitution of Eqs. (5)-(8) into Eq. (4) yields
\begin{equation}
\Sigma^r_{d_1}(\epsilon)=-\frac{i}{2}\Gamma_L+
\frac{(\epsilon-\epsilon_4+
\frac{i}{2}\Gamma_R)[t^2_{12}(\epsilon-\epsilon_3)+
t^2_{13}(\epsilon-\epsilon_2)]-(t^2_{12}t^2_{34}+t^2_{13}t^2_{24})
+2cos\phi
t_{12}t_{13}t_{24}t_{34}}{(\epsilon-\epsilon_2)(\epsilon-\epsilon_3)(\epsilon-\epsilon_4+
\frac{i}{2}\Gamma_R)-
t_{24}^2(\epsilon-\epsilon_3)-t_{34}^2(\epsilon-\epsilon_2)}.
\end{equation}
It can be seen from the above expression that the ${\bf AB}$
effect manifests itself in an additional  term  $2(cos\phi-1)
t_{12}t_{13}t_{24}t_{34}$ in the self-energy $\Sigma_{d_1}$ of
dot $1$ Green's function, the dominant contribution to the
spectral density of the structure.

 We now have to calculate the lesser Green's function
$G^<_{d_1d_1}$. It can be obtained from the Keldysh formula
\begin{equation}
G^<_{d_1d_1}=G^r_{d_1d_1}\Sigma^<_{d_1}G^a_{d_1d_1}=\Sigma^<_{d_1}|G^r_{d_1d_1}|^2,
\end{equation}
where
\begin{equation}
\Sigma^<_{d_1}=if_L(\epsilon)\Gamma_L+\sum\limits_{i=2,3}t^2_{1i}\tilde
G^<_{d_id_i}+t_{12}t_{13} e^{-i\phi}\tilde
G^<_{d_2d_3}+t_{12}t_{13} e^{i\phi}\tilde G^<_{d_3d_2}.
\end{equation}
The lesser Green's functions $\tilde G^<_{d_id_j}$ ($i,j=2,3$) can
be calculated from the following Keldysh equation
\begin{equation}
\tilde G^<_{d_id_j}=\sum\limits_{m,n=2,3}\tilde G^r_{d_id_m}
\tilde \Sigma^<_{d_md_n} \tilde G^r_{d_nd_j}, ~~~i,j=2,3
\end{equation}
where
\begin{equation}
\tilde \Sigma^<_{d_md_n}=\frac{if_R(\epsilon)
\Gamma_Rt_{m4}t_{n4}}{(\epsilon-\epsilon_4)^2+\Gamma_R^2/4}.
~~~m,n=2,3
\end{equation}
Here $f_R(\epsilon)=[1+e^{(\epsilon-\mu_R)/k_BT}]^{-1}$ is the
Fermi-Dirac distribution function of the right lead, with chemical
potential $\mu_R$. It is worth noticing that the above procedure
of calculating the various kinds of Green's functions can be
verified to give  the same results as the equation-of-motion
method \cite{Zubarev}. Combining Equations $(6)-(10)$ we find
\begin{equation}
G^<_{d_1d_1}=(if_L(\epsilon)\Gamma_L-2if_R(\epsilon)Im[\Sigma^r_{d_1}
+\frac{i}{2}\Gamma_L])|G^r_{d_1d_1}|^2.
\end{equation}
Notice that
\begin{equation}
G^r_{d_1d_1}-G^a_{d_1d_1}=(\frac{1}{G^a_{d_1d_1}}-\frac{1}{G^r_{d_1d_1}})|G^r_{d_1d_1}|^2
=2iIm[\Sigma^r_{d_1}]|G^r_{d_1d_1}|^2.
\end{equation}
Substituting Eqs. (14) and (15) into Eq. (2), one then obtains the
current entering the structure from the left lead  ,
\begin{equation}
J=J_L=\frac{2e}{h}\int d\epsilon[f_L(\epsilon)-f_R(\epsilon)]\cT
(\epsilon),
\end{equation}
in which
\begin{eqnarray}
\cT (\epsilon)&=&\frac{\Gamma_L(\epsilon) \Gamma_R(\epsilon)
}{\Gamma_L(\epsilon)+\Gamma_R(\epsilon)} \rho(\epsilon), \cr
 \rho(\epsilon)&=& -\frac{2({\Gamma_L(\epsilon)+\Gamma_R(\epsilon)})}{\Gamma_R(\epsilon)}
 Im[\Sigma^r_{d_1}(\epsilon)
+\frac{i}{2}\Gamma_L(\epsilon)]|\frac{1}{\epsilon-\epsilon_1-
\Sigma^r_{d_1}(\epsilon)}|^2
\end{eqnarray}
is the transmission probability for an electron passing through
the quantum dot square structure. Equations (16) and (17) are the
central results of this work. It can be seen that the transmission
probability and current are proportional to the dot-lead couplings
($\Gamma_L, \Gamma_R$)  and a generalized spectral density
function $\rho(\epsilon)$ of the system.  Note that the current
includes contributions from the upper arm path ($1\rightarrow
2\rightarrow 4$), and the bottom arm path (($1\rightarrow 3
\rightarrow 4$), which interfere with each other. At zero
temperature, the conductance in the linear regime limit $\mu_L
\approx \mu_R \rightarrow \epsilon_F$ becomes
\begin{equation}
\cG (\Phi)=\frac{2e^2}{h}\frac{\Gamma_L \Gamma_R
}{\Gamma_L+\Gamma_R} \rho(\epsilon_F),
\end{equation}
In obtaining Eq.(18) the wide-bandwidth approximation \cite{Meir}
has been used, i.e., the linewidths $\Gamma_L$ and $\Gamma_R$ are
constants independent of energy. From the expression of the
self-energy $(9)$ one can find that the linear conductance
$\cG(\Phi)$ is a periodic function of the magnetic flux $\Phi$,
with periodicity $\Phi_0=h/e$, and $\cG(\Phi)=\cG(-\Phi)$. These
observations are consistence with the general results of the
two-terminal setup of the ${\bf AB}$ ring \cite{Hackenbroich}.

 When quantum  dot $2$ ( or
3) is decoupled from the system, i.e., $t_{12}=t_{24}=0$ ( or
$t_{13}=t_{34}=0$), we recover the results for the coupled triple
quantum dot chain \cite{Zeng}
\begin{equation}
J=J_L=\frac{2e}{h}\int
d\epsilon[f_L(\epsilon)-f_R(\epsilon)]\cT_3(\epsilon),
\end{equation}
where the transmission probability is
\begin{equation}
\cT_3
(\epsilon)=\Gamma_L\Gamma_R|t_{1i}|^2|t_{i4}|^2\cA(\epsilon).~~~i=2
~~( or~~3)
\end{equation}
Here
\begin{eqnarray}
\cA(\epsilon)&=&\frac{[(\epsilon-\epsilon_4)^2+\Gamma_R^2/4]/\{[(\epsilon-\epsilon_i)
[(\epsilon-\epsilon_4)^2+\Gamma_R^2/4]-t_{i4}^2
(\epsilon-\epsilon_4)]^2+|t_R|^4\Gamma_R^2/4\}}
{(\epsilon-\epsilon_1-t_{1i}^2Re\cB(\epsilon)^2+(\Gamma_L-2t_{1i}^2
Im\cB(\epsilon)^2/4}, \cr
 i&=&2~~( or~~3)
\end{eqnarray}
where
\begin{equation}
\cB(\epsilon)=
(\epsilon-\epsilon_i-\frac{t_{i4}^2}{\epsilon-\epsilon_4+\frac{i}{2}\Gamma_R})^{-1}.~~~i=2
~~( or~~3)
\end{equation}
When the levels of the three dots are aligned, we find the
condition for complete transmission
  ($\cT_3=1$) to be
\begin{equation}
\frac{t_{1i}^2}{t_{i4}^2}=\frac{\Gamma_L}{\Gamma_R}. ~~~i=2 ~~(
or~~3)
\end{equation}
It can be expected that the criteria for complete resonant
transmission or maximum peak conductance plays an important role
in the formation of  conductance peaks with the conductance quanta
$2e^2/h$ for the square  quantum dot array.

\section{Linear conductance characteristics}
\subsection { Symmetric coupling between dots and leads}

It is well known that the two-terminal conductance  of an ${\bf
AB}$ ring is a periodic even function of the enclosed magnetic
flux $\Phi$ \cite{Hackenbroich}. The oscillation of the
conductance, with  periodicity $h/e$, results from the quantum
interference between the paths located in the upper and lower arms
of the ring. For electron tunneling through a single dot, the
conductance peak reaches  its maximum when $\Gamma_L=\Gamma_R$. In
the coupled double quantum dot structure with aligned energy
levels, the conductance peak reach a maximum value and splits into
a double peak with peak separation $2(t^2-\Gamma^2/4)^{1/2}$ when
the interdot coupling $t$ becomes larger than the width of the
peak $\Gamma$ in the symmetric dot-lead coupling case\cite{Zeng}.
For the three coupled dot string, the situation is more
complicated, though the condition for maximum conductance peak is
the simple Eq. (23) since the levels of the dots are aligned. We
expect that the conductance peak will split into three maxima when
the interdot coupling becomes larger than the width of the peak,
as in the double dot case.

For further comparison, we first calculate the linear conductance
of the quantum dot array in the absence of the magnetic flux
$\Phi$ and in the symmetric dot-lead coupling case
($\Gamma_L=\Gamma_R$). Throughout this paper,  energy and coupling
strengths are measured in units of $\Gamma_L=\Gamma$. The results
for aligned levels and various kinds of interdot coupling
configurations are shown in Fig. 2 (a), while Fig. 2 (b,c) are for
different arrangements of the dot levels.  When the four dot
levels are aligned,  three conductance peaks appear if
$t_{12}=t_{13}$ and $t_{24}=t_{34}$.  Then the conductance
possesses the same three resonant peak structure as the three dot
string. One can also find that the criteria for complete resonant
transmission for the three-dot systems manifests itself in our
four dot structure. This is not strange since the conductance in
our case is simply the superposition of the contributions from
 the upper and lower three-dot arms. When
$t_{12}=t_{34}<t_{13}=t_{24}$, the situation becomes different. In
this case, one can consider the  dot structure as a system of two
coupled quantum dot dimers. One observes in the conductance
spectra four peaks, arranged in two groups separated by a
conductance gap. The separation between the two group centers is
mainly determined by the larger interdot coupling, and the peak
separation of each group by the smaller one. The peak and group
center separations have trivial dependence on the dot-lead
couplings $\Gamma_L$ and $\Gamma_R$ in the coupled quantum dimers
case. The conductance gap is introduced by the larger coupling
strength between the two  dimers and can not be lifted  by a
magnetic flux, contrary to the case of the Fano conductance dip
discussed below. The gap can be broadened by increasing the
coupling strength(s) $t_{12}$( and/or $t_{24}$) between the two
dimers.

Figures 2 (b,c), displaying cases of mismatched levels,  show more
interesting conductance spectra. When
$t_{12}=t_{13}=t_{24}=t_{34}$ a striking novel conductance dip
($\cG=0$) appears {\it halfway between the energy levels of dots
$2$ and $3$} (Fig. 2(c) and dashed line in 2(b)) if these levels
are different. When $t_{12}=t_{24}<<t_{13}=t_{34}$  the
conductance dip is pinned at the energy level of the quantum dot
with small couplings to the system (solid line in Fig. 2 (b)). The
middle peak appears split into two asymmetric peaks. The asymmetry
between the two split peaks depends on the mismatch between the
energy levels of dots 2 and 3, as well as the arrangement of the
levels of other dots 1 and 4. Novel five spikes ( the dashed line
in Fig. 2 (b)) in the conductance can be seen if the mismatch
between the levels of dot 2 and 3 is large enough. The dip results
from a Fano interference\cite{Fano} between two distinct current
paths,  one being direct nonresonant and the other nondirect
resonant\cite{Nockel,Kim}. The conductance can be generally
written\cite{Nockel} as
$\cG=\cG_{non}\frac{(\tilde{\epsilon}+q)^2}{\tilde{\epsilon}^2+1}$,
where $\cG_{non}$ is the nonresonant conductance,
$\tilde{\epsilon}=\epsilon-\epsilon_{reso}$, and $q$ is the 'Fano
parameter' assessing the asymmetry of the lineshape. Two limiting
cases exist [Ref. 17(b)]: (a) $q \rightarrow \infty$, with a
dominating resonant transmission resulting in Breit-Wigner
resonances and,  (b) $q \rightarrow 0$, with a dominating
nonresonant transmission leading to asymmetric dips. Since here
$t_{12}=t_{13}=t_{24}=t_{34}$,  one can consider either the arm
consisting of dot 2 or
 3 as the resonant (nonresonant) path. {\it This special symmetry
 locates  the Fano-type conductance dip  at neither the level
 of  dot 2 nor that of
dot 3, but halfway between them}.  On the other hand, when
$t_{12}=t_{13}=t_{24}=t_{34}$, we find from Eqs. (9), (17) and
(18) that $\cG \propto \frac{2\epsilon_F-\epsilon_2-\epsilon_3}
{(\epsilon_F-\epsilon_2)(\epsilon_F-\epsilon_3)}$. It is evident
that two resonance peaks exist when $\epsilon_F=\epsilon_2$ and $
\epsilon_F=\epsilon_3$, while $\cG$=0 when
$\epsilon_F=(\epsilon_2+\epsilon_3)/2$, which is just the position
of the Fano conductance dip. For large mismatch between  energy
levels of dots 2 and 3, Fano interference between the two paths
develops  both a conductance dip and a peak halfway between these
levels $\epsilon_2$ and $\epsilon_3$. One can then observe a five
peak structure in the conductance.
 When the levels of dots
2 and 3 are matched, the resonant peak structure of the dot
systems behaves to a large extent the same way as the coupled
three-dot system \cite{Zeng}.

   Now we turn to study the effect of the  magnetic flux $\Phi$
    on the conductance spectra of our square dot
   system. Figure 3 exhibits clearly the {\bf AB} oscillations of the
    conductance as a function of the magnetic flux $\phi$
   within a period at $\epsilon_F=0.1 \Gamma$.
   When the two arms of the dot ring are symmetric, i.e., the dot
   levels and interdot couplings are arranged to be the same, the
   conductance $\cG$ vanishes at $\phi=\pi$ (Fig. 3(a) and the solid
   line in Fig. 3(c)). In the case of asymmetric arms, the
   conductance is nonzero within the whole period. The
   vanishing of $\cG$ at $\phi=\pi$ results from the complete
   destructive interference between the two symmetric arms, which
   will be analyzed  in more detail below.  If the two arms are not
   symmetric, complete destructive interference is never
   achievable, and  the conductance is always nonzero.
    The more asymmetry between the two arms, the
   larger the conductance near $\phi=\pi$. To get a clearer
   insight into the influence of a magnetic flux on $\cG$,
    we give in Figs. 4 and 5 the conductance of
   the structure as a function of the Fermi energy for different
   values of
   magnetic flux within half a period ($\phi=0, \pi/4, 3\pi/4,
   \pi$),
   and different configurations of interdot couplings and dot levels.
   The dependence of
   the conductance on the magnetic flux $\Phi$ is apparent. The
   magnitude of the ${\bf AB}$  oscillation
   strongly depends on the matching condition of the dot levels
   and the configuration of the interdot couplings. It  also
    differs for different pinning positions
   of the Fermi energy at the leads. The presence
   of a magnetic flux can lead to the splitting (Fig. 5(b)) and combination
   (Fig. 4(c) and Figs. 5 (a,c,d))
   of conductance peak(s) since it changes the interference between the two
   current paths, as determined by
    term  $2(cos\phi-1)
   t_{12}t_{13}t_{24}t_{34}$ in Eq. (9).
   Another striking effect induced by the
   magnetic flux is that the central conductance peak, if present (Fig. 4),
   becomes a conductance dip when the levels of dots $2$ and $3$ are aligned,
   while the Fano-type dip  appearing in the case of mismatched levels  is
   lifted (Fig. 5), contrary to the usual conductance gap which survives in the presence of
   a magnetic flux (Fig. 4 (c)).
    This feature is
   typically a result of the modified Fano phase interference by the
   magnetic flux. When $\phi=\pi$, the conductance is
   depressed everywhere if $\epsilon_2=\epsilon_3$ and
   $t_{12}t_{24}=t_{13}t_{34}$, as shown in Fig. 4  and Fig. 5 (b).
   The overall depression of the conductance results from the
   complete destructive interference between the upper-arm and lower-arm
   current paths.  When $\epsilon_2=\epsilon_3=\epsilon$ and
   $t_{12}t_{24}=t_{13}t_{34}$,
   one finds $\cG \propto 2(1+cos\phi)t^2_{12}t_{24}^2$, which
   implies zero conductance for any value of the Fermi energy  if $\phi=\pi$.
   We can also analyze it from the interference pattern of the two current
   paths containing either  dot 2 or 3.
   The respective transmission amplitudes $T_{2}=t_{12}t_{24}/(\epsilon_F-\epsilon_2+i0^+)$
   and $T_{3}=t_{13}t_{34}e^{i\phi}/(\epsilon_F-\epsilon_3+i0^+)$
    equal $T/(\epsilon_F-\epsilon+i0^+)$ and
    $Te^{i\phi}/(\epsilon_F-\epsilon+i0^+)$, respectively, when
   $t_{12}t_{24}=t_{13}t_{34}=T$ and $\epsilon_2=\epsilon_3=\epsilon$.
    Then the total transmission
   probability $\cT=|T_2+T_3|^2$ for electron tunneling through the dot structure
   is proportional to $2T^2(1+\cos \phi)\delta (\epsilon_F-\epsilon)$,
   which is zero when $\phi=\pi$.

\subsection { Asymmetric coupling between dots and leads}

In the preceding subsection we have reported in detail the linear
conductance spectra in the absence and  presence of a magnetic
flux, in the limit of symmetric dot-lead couplings. There we saw
symmetrically-located conductance peaks if the dot levels are
aligned, which is a consequence of the particle-hole symmetry. In
this subsection, we want to study how the conductance is modified
if the dot-lead couplings are asymmetric.

We model the asymmetry between the dot-lead coupling $\Gamma_L$
and $ \Gamma_R$ by setting $\Gamma_R=\Gamma_L/2$. Figures 6 and 7
present the comparison of conductance spectra in the cases of
symmetric (thin lines) and asymmetric (thick lines) dot-lead
couplings, with structure parameters the same as in Fig. 4 and 5,
respectively. A surprising feature in these figures is that the
asymmetry between dot-lead couplings can remove the symmetry of
the conductance spectra around some energy when the energy levels
of dots are not aligned (Figs. 7 (c,d)), while it can not if they
are aligned (Fig. 6). Also depending on the value of the magnetic
flux $\phi$, the conductance in the case of asymmetric dot-lead
couplings is lifted (Figs. 6 (a,c), $\phi=3 \pi/4$) or suppressed
(Figs. 6 (a,c), $\phi=\pi/4$ and Fig. 7 (a) $\phi=\pi/4, 3\pi/4$),
as compared to the symmetric case. We also observe another
interesting phenomenon. The conductance changes in a nonuniform
way when the dot-lead couplings become asymmetric. This is
reflected in the observation that some conductance peaks are
lifted while the others are suppressed (Fig. 6 (b) and  Figs. 7
(c,d)). In addition, we find that the asymmetric dot-lead
couplings would introduce a splitting of the conductance peak
(Fig. 6 (b), Figs. 7 (c,d)). It is worth mentioning that the
position of the conductance peaks remain nearly unshifted when the
dot-lead couplings become asymmetric.

 These features can be understood for linear conductance from the
analysis of the generalized spectral function $\rho(\epsilon)$ in
Equation (17), which depends on  the dot-lead couplings $\Gamma_L$
and $\Gamma_R$.  In the case of a single quantum dot,a difference
in the value of the dot-lead couplings $\Gamma_L$ and $\Gamma_R$
does not change the position of the resonant transmission or
conductance peak, but does alter its height. In our case, the
dependence on $\Gamma_L$ and $\Gamma_R$ of the spectral function
or conductance is much more complicated. However, it appears that
a difference between $\Gamma_L$ and $\Gamma_R$ does also modify
the weight of the peaks in the generalized spectral function
$\rho(\epsilon)$ in a complicated way dependent on the applied
magnetic flux. The position of the conductance peaks is kept
intact by the difference between $\Gamma_L$ and $\Gamma_R$.

\section{Conclusions}

To sum up, we have investigated the transport properties in the
linear regime, of a two-terminal quantum dot square enclosing a
magnetic flux. Although our results are not directly applicable to
more complex arrays involving more conducting paths and dots , we
believe that our study will provide insight into the physics of
interference that such arrays will exhibit.  Our main results may
be summarized as follows: (1) The ${\bf AB}$ oscillations of the
conductance strongly depends on the configurations of dot levels
and interdot and dot-lead couplings, along with the position of
the Fermi energy of the leads; (2) a striking conductance dip is
developed due to the Fano interference when the levels of the
intermediate dots are mismatched, and a novel five peak structure
is observed for large mismatch; (3) the magnetic flux and an
asymmetric dot-lead coupling can induce the splitting and
combination of the conductance peak(s); (4) the interplay of a
magnetic flux and asymmetry between dot-lead coupling can lead to
interesting ${\bf AB}$ oscillations of the linear conductance. If
the electron-electron interactions are taken into consideration,
further structure in the conductance may be expected due to the
presence of an additional energy scale.  Results in the regime in
which such energy scale is relevant will be reported elsewhere.

\begin{center}
{\bf ACKNOWLEDGMENT}
\end{center}
This work was supported by a C\'atedra Presidencial en Ciencias,
FONDECYT 1990425 (Chile) and NSF grant No. 53112-0810 of Hunan
Normal University (China). Discussions with P. Orellana are
acknowledged.


\newpage
\begin{figure}
\caption{ Schematic plot of a 2$\times$2 quantum dot structure
enclosing a magnetic flux $\Phi$ with two terminals.}
\end{figure}

\begin{figure}
\caption{Conductance $\cG$ as a function of Fermi energy
$\epsilon_F$ in the case of symmetric dot-lead couplings
($\Gamma_L=\Gamma_R$) and in the absence of magnetic flux when the
levels of the four dots are aligned (a) and not aligned (b, c).}
\end{figure}

\begin{figure}
\caption{{\bf AB} oscillations of the conductance $\cG$ as a
function of magnetic flux $\phi$ in the case of symmetric dot-lead
couplings ($\Gamma_L=\Gamma_R$). The structure parameters are the
same as in Fig. 2 and the Fermi energy $\epsilon_F$ is set to be
$0.1\Gamma$.}
\end{figure}

\begin{figure}
\caption{Evolution of the conductance $\cG$ with the Fermi energy
$\epsilon_F$ for the case of symmetric dot-lead couplings
($\Gamma_L=\Gamma_R$) and the levels of the four dots aligned, and
$\phi=0$ (solid), $\phi=\pi/4$ (dashed), $\phi=3\pi/4$ (dotted)
and $\phi=\pi$ (dash-dotted).}
\end{figure}

\begin{figure}
\caption{Same as Fig. 4, but with the levels of the four dots not
aligned.}
\end{figure}

\begin{figure}
\caption{Conductance $\cG$ in the cases of symmetric dot-lead
couplings ($\Gamma_L=\Gamma_R$, thin line) and asymmetric dot-lead
couplings ($\Gamma_R=\Gamma_L/2$, thick line) for aligned four dot
levels. The solid and dotted lines correspond to $\phi=\pi/4$ and
$3\pi/4$, respectively. }
\end{figure}

\begin{figure}
\caption{Same as Fig. 6 but for non-aligned levels.}
\end{figure}


\begin {references}
\bibitem {AB} Y. Aharonov and D. Bohm, Phys. Rev. {\bf 115}, 485
(1959).
\bibitem {Webb} S. Washburn and R. A. Webb, Adv. Phys. {\bf 35}, 375
(1986).
\bibitem {Beenakker} C. W. J. Beenakker, H. van Houten, and A. A.
M. Staring, Phys. Rev. B {\bf 44}, 1657 (1991), and references
therein.
\bibitem {Yacoby} A. Yacoby, M. Heiblum, D. Mahalu, and H. Shtrikman,
Phys. Rev. Lett.{\bf 74}, 4047 (1995); R. Schuster, E. Buks, M.
Heiblum, D. Mahalu, and H. Shtrikman, Nature {\bf 385}, 417
(1997).
\bibitem {Hackenbroich} G. Hackenbroich, Phys. Rep. {\bf 343}, 463
(2001), and references therein.
\bibitem {Kouwenhoven} For a review, see L. P. Kouwenhoven, C. M. Marcus, P. L. McEuen,
S. Tarucha, R. M. Westervelt, and N. S. Wingreen, in {\it
Mesoscopic Electronic Transport}, Edited by L. L. Sohn, L. P.
Kouwenhoven, and G. Sch{\"o}n, (Kluwer, Series E 345, 1997), P
105-214.
\bibitem {Averin} D. V. Averin and Yu. V. Nazarov, Phys. Rev. Lett. {\bf 65}, 2446
(1990).
\bibitem {Datta} S. Datta, {\it Electronic Transport in Mesoscopic
Systems}  (Cambridge University Press, 1995), P246-273.
\bibitem {Haug} L. P. Kouwenhoven, F. W. J. Hekking, B. J.
van Wees, C. J. P. M. Harmans, C. E. Timmering, and C. T. Foxon,
Phys. Rev. Lett. {\bf 65}, 361 (1990); R. J. Haug, J. M. Hong and
K. Y. Lee, Surf. Sci. {\bf 263}, 415 (1992).
\bibitem {Stafford} C. A.
Stafford and S. Das Sarma, Phys. Rev. Lett. {\bf 72}, 3590 (1994);
G. Klimeck, G. Chen, and S. Datta, Phys. Rev. B {\bf 50}, 2316
(1994); F. R. Waugh, M. J. Berry, D. J. Mar, R. M. Westervelt, K.
L. Kampman, and A. C. Gossard, Phy. Rev. Lett. {\bf 75}, 707
(1995).
\bibitem {Oosterkamp}  T. H. Oosterkamp, T. Fujisawa, W.
G. van der Wiel, K. Ishibashi, R. V. Hijman, S. Tarucha,  and L.
P. Kouwenhoven, Nature {\bf 395}, 873 (1998); R. H. Blick, D.
Pfannkuche, R. J. Haug, K.v. klitzing, and K. Eberl, Phys. Rev.
Lett. {\bf 80}, 4032 (1998); R. H. Blick, D. W. van der Weide, R.
J. Haug, and K. Eberl, ibid. {\bf 81}, 689 (1998).
\bibitem {Yu} Z. Yu, T. Heinzel, and A. T. Johnson, Phys. Rev. B
{\bf 55}, 13697 (1997).
\bibitem {Meir} Y. Meir  and N. S. Wingreen, Phys. Rev. Lett. {\bf 68}, 2512 (1992);
 A. -P. Jauho, N. S. Wingreen, and Y. Meir, Phys. Rev. B {\bf 50}, 5528 (1994).
\bibitem {Zeng} Z. Y. Zeng, F. Claro, and W. Yan, e-print,
Cond-matt/0105194.
\bibitem {Zubarev} D. N. Zubarev, Usp. Fiz. Nauk {\bf 71}, 71
(1960) [Sov. Phys. Usp. {\bf 3}, 320 (1960)].
\bibitem {Fano} U. Fano, Phys. Rev. {\bf 124}, 1866 (1961).
\bibitem {Nockel} (a) J. U. N\"ockel and A.
Douglas Stone, Phys. Rev. B {\bf 50}, 17415 (1994); (b)
 J.G\"ores, D. Goldhaber-Gorden, S. Heemeyer, and M. A. Kastner,
Phys. Rev. {\bf 62}, 2188 (2000); (c) A. A. Clerk, X. Waintel, and
P. W. Brouwer, Phys. Rev. Lett. {\bf 86}, 4637 (2001).
\bibitem {Kim} T. S.
Kim and S. Hershfield, Phys. Rev. B {\bf 63}, 245326 (2001);
\end{references}
\end{document}